\documentclass[proceedings,eqalign]{JHEP2} % 10pt is ignored!


% ...               

\def\@oddhead{RAts} 

\usepackage{epsfig}                   % please use epsfig for figures
% ...                                   
\def\I{{\cal I}}
\def\P{{\rm P}}
\def\Pi{{\rm P}}

\def\NPi{{\rm NP}}

\def\e{\epsilon}
\def\eps{\epsilon}
\def\Tr{\, {\rm Tr}}
\def\eqn{}

\def\kapss{\left( { \kappa \over 2 } \right)^4 }

\def\Mtree{M^{\rm tree}}

\def\npb#1#2#3{{\it Nucl.\ Phys.\ } {\bf  B#1}:#3 (#2)}
\def\plb#1#2#3{{\it Phys.\ Lett.\ } {\bf  B#1}:#3 (#2)}
\def\pr#1#2#3{{\it Phys.\ Rev.\ } {\bf  #1}:#3 (#2)}
\def\cqg#1#2#3{{\it Class.\ and Quant.\ Grav.} {\bf  #1}:#3 (#2)}
\def\prl#1#2#3{{\it Phys.\ Rev.\ Lett. } {\bf  #1}:#3 (#2)}
\def\JHEP#1#2#3{{\it J.\ High\ Ener.\ Phys.\  } {\bf  #1}:#3 (#2)}
\def\ijmp#1#2#3{{\it Int.\ J.\ Mod.\ Phys.\ } {\bf A #1}:#3 (#2)}
\def\jmp#1#2#3{{\it  J. Math.\ Phys.\ } {\bf  #1}:#3 (#2)}
\newskip\humongous \humongous=0pt plus 1000pt minus 100pt
\def\caja{\mathsurround=0pt}
\def\eqalign#1{\,\vcenter{\openup1\jot \caja
       \ialign{\strut \hfil$\displaystyle{##}$&$
        \displaystyle{{}##}$\hfil\crcr#1\crcr}}\,}
\newif\ifdtup

\conference{Non-Perturbative Quantum Effects 2000}
\title{Counterterms in Supergravity}
\author{ 
{\bf \hskip -1.0 truecm
Z. Bern, D. Dunbar, L. Dixon, B. Julia, M. Perelstien, J. Rozowsky,
D. Seminara and M. Trigante} 
\\
Talk presented by D. Dunbar 
\\
University of Wales Swansea \\
        E-mail: \email{d.c.dunbar@swan.ac.uk}
\thanks{Research supported by TMR contract FMRX-CT92-0012}}

\abstract{We  examine the ultraviolet behaviour of supergravity theories
as a function of dimension and number of supercharges.
We do so by
the computation of one and two-loop physical on-shell four point amplitudes.

For maximal supergravity, our computations prove the non-renomalisability of
supergravity for $D \geq 6$ (including the maximal $D=11$ case) and give strong
evidence for the existance of a five-loop counterterm in $D=4$. For
type~I supergravity our results indicate similar patterns.

We shall also explore a remarkable relationship between gravity
amplitudes and those of Yang-Mills theories. In many ways gravity
calculations discover features which relate to the equivalent
Yang-Mills features by a squaring proceedure. 
}

\begin{document}

\section{Motivation}

Supergravity is one of the key theories in understanding quantum gravity. 
In itself, it is almost certainly not a renomalisable theory so must
appear as the low energy limit of another theory such as M-theory. 
However, supergravity and variations thereof will be the effective theories
which describe quantum gravity at energies less than the (colossally large)
Planck scale.

We shall study the ultra-violet behavior of supergravity theories for
two reasons: firstly we wish to prove the conjectured bad behavior of
these theories: secondly we hope to understand some features of the
physics at the Planck scale. Adding counterterms is a well defined,
but unpredictive at the Planck scale, way to regulate a theory. A
physical regulator should in some sense provide the same regulation but within
a predictive context. The symmetries and structure of the physical theory 
might well find themselves mirrored in the counterterm structure. 

One of the themes of this TMR network has been the use of
integrability in understanding two dimensional field
theories. Unfortunately the enormous success in two dimensions has not,
yet, continued to higher dimensions.  However, the techniques we use
have some formal similarities. We can construct $S$-matrix elements from
the analytical nature of the amplitudes. However we fall a long way
short of the exact $S$-matrices found in two dimensions.  Nonetheless,
we are able to construct enough of the $S$-matrix elements to determine
large amounts of the ultra-violet structure of supergravity theories in
dimension four or greater. In constructing amplitudes we attempt to
use any and all information regarding the amplitude. Supersymmetry
is one useful tool and in gravity theories with extended
supersymmetry we can make more progress.  For the maximal supergravity
theory we can prove its non-renormalisability in $D=11$ and can
conjecture the behavior in $D=4$.

\section{Technology}

Our philosophy is to evaluate the physical, onshell $S$-matrix from
it's analytic properties.  As far as possible we shall only consider
on-shell objects.

\EPSFIGURE[l]{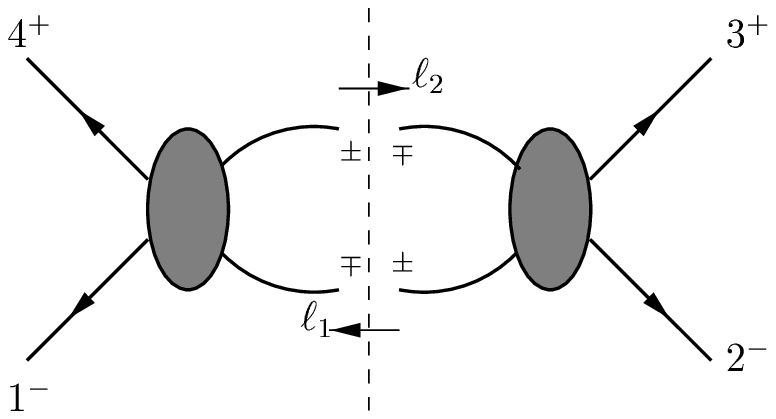,width=3.5 truecm}{}

A key property of the $S$-matrix is {\it unitarity}.  Also within
dimensional regularisation the amplitude is analytic in the dimension. 
The optical theorem, a consequence of unitarity states
$$
2 Im T =T^{\dagger}T
$$
In perturbation theory comparing both sides order by order relates,
for example, the imaginary part of a one-loop amplitude to the product
of tree amplitudes.  In practical terms the imaginary part of a one loop
amplitude is just the coefficient of a logarithm (or di- or polylogarithms)
since,
$$
\ln(s_{iJ}) = \ln(|s_{ij}|)+i\pi \Theta(s_{ij}) 
$$
where $s_{ij}$ is one of the momentum invariants. 
Naively, the optical theorem only determines part of the one-loop
amplitude since the amplitude may contain rational functions
$f(s_{ij})$ which have no imaginary part. However we can, by using
dimensional regularisation, determine these rational parts also. 
Within dimensional regularisation the one-loop
amplitude has a momentum weight of
$-2\eps$ (since $d^Dp \longrightarrow d^{D-2\eps}$)  This implies that
the rational functions must be replaced by terms such as
$f'(s_{ij})(s_{ij})^{-2\eps}$.
Since
$$
(s_{ij})^{-2\eps} =1-2\eps \ln(s_{ij})
$$
the amplitude will pick up imaginary parts at $O(\eps)$.  
We thus deduce

{\it knowledge of the cuts to all powers in $\eps$ will enable us to determine the amplitude}

Of course this can be a painful computational burden  in many circumstances. 
(Although in practice it is not always necessary to determine the cuts to
all orders in $\eps$.) 
To see how this works consider the cut in a one-loop amplitude.
Then the optical theorem states\cite{Cutting},
$$
\eqalign{
 & {\rm Disc}~M^{1-loop}(1,2,3,4)
\Bigr|_{s-cut}
=
\cr
 & i \int\sum_{ {\rm internal}\atop {\rm particles}}
\,\Mtree(-\ell_1^s,1,2,\ell_2^{s'})\, 
%\cr
%& \null\hskip 1.5 truecm \times 
\Mtree(-\ell_2^{s'}, 3,4,\ell_1^s )
\cr}
\eqn
$$
where the integral is over on-shell $\ell_i$. 
We must use this carefully within dimensional regularisation
if we wish to determine the LHS correctly to all orders in $\eps$.  
The RHS contains tree amplitudes. Normally we do not regard these are 
depending upon $\eps$ however the momenta $\ell_i$ must match the
loop momenta in the LHS. These are in $D-2\eps$ dimensions so that the
tree amplitudes should have the momenta $\ell_i$ in $D-2\eps$ and the
others in  $D$ dimensions.

The analysis here is naturally merely indicative and the reader must
be referred to elsewhere for the details of how this works and how it
may be applied.

The optical theorem thus stated is a key ingredient to our 
calculational programme however it is not the only important 
input we also use some or all of the following features

$\bullet$ Amplitudes may in principle be calculated using Feynman diagrams.
This allows us to restrict the ``function'' space an amplitude may lie in.

$\bullet$ Supersymmetric theories generally have simpler amplitudes
which can be easier to calculate

$\bullet$ Field theory amplitudes may be calculated as the low energy limit
of string theory amplitudes~\cite{StringBased}.

$\bullet$ Amplitudes should have factorisation and collinear singularities
when momentum invariants have specific values~\cite{BCM}. 

\section{ $N=8$ Supergravity Amplitudes}

Maximal supergravity \cite{ExtendedSugra,CJS} 
is a fascinating theory whose ultraviolet behavior
is suspected but until the last few years has defied definite calculation.
We shall attempt to determine this.

The one-loop amplitude was calculated many years ago by Green
and Schwarz and Brink \cite{GSB} to be
$$
{\cal M}_4^{\rm one-loop}  = \Bigl({\kappa \over 2} \Bigr)^2 \, stu \, 
 M_4^{\rm tree}
 \Bigl( \I_4(s, t)+ \hbox{ 2 perms.}
\Bigr)
$$ 
where $\I_4(s, t)$ is the $D$-dimensional scalar box integral (which
may be easily evaluated).  The one-loop amplitude is infinite in $D=8$
but not in other dimensions- on-shell. To determine the behavior in other dimensions we must go beyond a one-loop calculation.

Using the technology described previously we
have obtained a remarkably simple result for the final form for the 
two-loop four graviton 
amplitude,~\cite{BDDPR}
$$
\eqalign{ 
&{\cal M}_4  = 
 \Bigl({\kappa \over 2} \Bigr)^6 \, stu \, 
 M_4^{\rm tree}
 \Bigl(s^2 \, \I_4^{\P}(s, t) 
+ s^2 \, \I_4^{\P}(s, u)  \cr
&  \null 
\hskip 1.0 truecm 
+ s^2 \, \I_4^{\NPi}(s, t)
+ s^2 \, \I_4^{\NPi}(s, u) 
+ \hbox{cyclic} \Bigr)  \cr}
\eqn
$$
where $\I_4^{\Pi}$ and $\I_4^{\NPi}$
are two-loop scalar box integrals. They are the planar and non-planar box respectively.  This amplitude has ultra-violet infinities in all dimension $D>6$.
In particular there is a definite divergence in the maximal dimension $D=11$.

The two-loop ultraviolet divergences for $N=8$ supergravity in $D=7$,
9 and 11, 
are
$$
\eqalign{
{\cal M}_4^{D=7}\vert_{\rm pole} 
 &= 
 {1\over2\e\ (4\pi)^7} {\pi\over3} (s^2+t^2+u^2) \, 
\times {\cal F}
%\left( {\kappa \over 2}\right)^6 
%\times stu M_4^{\rm tree} 
, \cr 
{\cal M}_4^{D=9}\vert_{\rm pole} 
 &=  {1\over4\e\ (4\pi)^9} {-13\pi\over9072} (s^2+t^2+u^2)^2 \, 
\times 
{\cal F}
%\left( {\kappa \over 2}\right)^6 \times stu M_4^{\rm tree} \, , 
\cr 
{\cal M}_4^{D=11}\vert_{\rm pole} 
 &=  {1\over48\e\ (4\pi)^{11}} \times
  \cr
 {\pi\over5791500} 
\Bigl( & 438 (s^6+t^6+u^6) - 53 s^2 t^2 u^2 \Bigr) \,
\times 
{\cal F}
%\left( {\kappa \over 2}\right)^6 \times stu M_4^{\rm tree} \, . \cr 
\cr}\eqn
%\label{gravtwolooppoles}
$$
where ${\cal F}=({\kappa/2})^6 \times stu M_4^{\rm tree}$.
There are no sub-divergences because one-loop divergences are absent in
odd dimensions when using dimensional regularisation.  
For even dimensions
$$
\eqalign{
{\cal M}_4^{D=8}\vert_{\rm pole} 
 &=  {1\over2 \ (4\pi)^{8}} \times \cr
&\Bigl( - {1\over 24 \, \e^2} 
                  + {1\over 144\e} \Bigr) \bigl(s^3 + t^3 + u^3 \bigr)
 \times 
{\cal F} \cr
{\cal M}_4^{D=10}\vert_{\rm pole} 
 &=  {1\over12\e\ (4\pi)^{10}} { - 13 \over 25920 } \times \cr
& \, stu \, 
  \bigl( s^2+t^2+u^2 \bigr) \times 
{\cal F}
\cr}\eqn
$$
The $1/\eps^2$ pole in $D=8$ is {\it precisely} that need to cancel
the $1/\eps^2$ pole obtained when the one-loop counterterm is used used
to calculate to two loops. The $1/\eps$ pole shows how the expected
non-predictive nature of renormalisation occurs - new terms must be
added to the Lagrangian order by order.

In all cases, for four graviton external states, the linearized
counterterms take the form of derivatives acting on
$$
\eqalign{
t_8 t_8 R^4 &\equiv
t_8^{\mu_1\mu_2\cdots \mu_8}\, 
t_8^{\nu_1\nu_2\cdots \nu_8} \, 
R_{\mu_1\mu_2 \nu_1 \nu_2} \, 
\cr
\times & 
R_{\mu_3\mu_4 \nu_3 \nu_4} \,
R_{\mu_5\mu_6 \nu_5 \nu_6} \,
R_{\mu_7\mu_8 \nu_7 \nu_8}  \,,
\cr}
\eqn
%\label{Rfourterms}
$$
plus the appropriate $N=8$ completion \cite{DS}.
-which also appears as the one-loop counterterm for $N=8$ 
supergravity in $D=8$.
This particular tensor is well known from 
string theory amplitudes \cite{GSW}, appears in the string effective action 
\cite{R4} and
is one of the higher dimensional analogs of the Bel-Robinson 
tensor \cite{BelRob}. It is consistent with $N=8$ supersymmetry which
may not allow other possibilities.

For example the $D=11$ counterterm is a linear combination of the two tensors
$$
\eqalign{
T^A&= t_{8} t_{8}\cdot
\partial_{\alpha\gamma\eta}R \partial^{\alpha\gamma\eta}R
\partial_{\beta\delta\rho}R \partial^{\beta\delta\rho}R
\cr
T^B&= t_{8} t_{8}\cdot
\partial_{\alpha\gamma\eta}R \partial^{\alpha\delta\eta}R
\partial_{\beta\gamma\rho}R \partial^{\beta\delta\rho}R
\cr}
$$
In each case the indices on the curvatures are contracted with the
$t_8$ tensors and the indices on the derivative are contracted with
each other.  The $D=11$ counterterm is 
$$
=  - {1\over 48\e\ (4\pi)^{11}} \times
 {\pi\over5791500}  \biggl( { 2575 \over 12} T^A+ {53 \over 6} T^B \biggr)
$$

\section{Higher-loop conjecture}

To determine the behavior for $D\leq 6$ we need to go beyond two
loops. As yet this remains a very challenging calculation.  In order
to specify the precise form of the conjecture at $L$ loops one would
need to investigate cuts with up to $(L+1)$ intermediate particles.
Nevertheless, some of the integral coefficients and numerators can be
obtained from the two-particle cuts.  If we assume these pieces
of the amplitude are representation we can conjecture the ultra-violet
structure. Of course in the absence of definite calculations this
remains very much a guess - however a guess which we expect will prove
correct.

By examine some of the cuts we can identify the most divergent pieces as

$$
\int (d^D p)^L {(p^2)^{2(L-2)} \over (p^2)^{3L +1}} \,. \eqn
$$
This integral will be finite when
$$
D < {10 \over L} + 2 \,,  \hskip 2 cm (L > 1) \,
\eqn
$$

The results of this analysis are summarized in
table~1.  In particular, in $D=4$ no three-loop
divergence appears - contrary to expectations from a superspace
analysis~\cite{HST,HoweStelle} 
- and the first $R^4$-type counterterm occurs at five loops.  
The divergence will have the same kinematical structure as the $D=7$
two-loop divergence , but with a different
non-vanishing numerical coefficient.

\begin{table}[ht]
\vskip .5 cm
\hskip -0.5 truecm
\hbox{
\def\tend{\cr \noalign{\hrule}}
% some compact definitions
\def\t#1{\tilde{#1}}
\def\tw{\theta_W}
\vbox{\offinterlineskip
{
\hrule
%\hrule
\halign{
        &\vrule#
%        & \vrule#
        &\strut \hfil #\hfil\quad\vrule
        &\hfil\strut#\hfil\quad\vrule
        &\hfil\strut#\hfil\quad\vrule
        & \hfil\strut # \hfil \vrule
        \cr
height13pt  &{ Dimension }  &{ Loop }   &{ Degree}    & { Counterterm} \tend
height12pt  & $8$ 
& 1 & log. &  $R^4$ \tend
height12pt  & $7$
& 2 &  log. & $\partial^4 R^4$ \tend
height12pt  & $6$
& 3 & quad. & $\partial^6 R^4$ \tend
height12pt  & $5$
& 4 & quad. & $\partial^6 R^4$ \tend
height12pt   & $4 $
& 5  &  log. & $\partial^4 R^4$ \tend
}
}
%\hrule
}
}
\nobreak
\caption[]{
%\label{CountertermsTableGR} 
\small The relationship between dimensionality and the number of loops 
at which the {\it first} ultraviolet divergence should occur in the 
$N=8$ supergravity four-point amplitude.  The form of the associated 
counterterm assumes the use of dimensional regularisation.
\smallskip}
\end{table}

\section{Non-maximal Supergravity}

It is interesting to compare the structures found between types II and
type I supergravity (and their lower dimensional descendants.)  We have
examined the one-loop structures for dimensions $4 \leq D \leq 10$
\cite{DJST}
however here we shall restrict presentation to the features of the
$D=8$ case. 

In $D=8$ power counting indicates that the counterterms will be of the form 
$R^4$.  There are seven independent $R^4$ tensors \cite{Fulling}
(in $ D < 8$ these are no longer independent.) 
\footnote{The Riemann tensor is
undistinguished from the Weyl tensor in our $R^4$ terms for on-shell
four point amplitudes.} 
$$
\eqalign{ T_1 =& ( R_{p,q,r,s}R_{p,q,r,s})^2 \cr T_2 =& (
R_{p,q,r,s}R_{p,q,r,t})( R_{p',q',r',s}R_{p',q',r',t}) \cr T_3 =&
R_{p,q,r,s}R_{p,q,t,u}R_{t,u,v,w}R_{r,s,v,w} \cr T_4 =&
R_{p,q,r,s}R_{p,q,t,u}R_{r,t,v,w}R_{s,u,v,w} \cr T_5 =&
R_{p,q,r,s}R_{p,q,t,u}R_{r,v,t,w}R_{s,v,u,w} \cr T_6 =&
R_{p,q,r,s}R_{p,t,r,u}R_{t,v,u,w}R_{q,v,s,w} \cr T_7=&
R_{p,q,r,s}R_{p,t,r,u}R_{t,v,q,w}R_{u,v,s,w} \cr} \eqn
$$ 

On shell the combination
$$
-{ T_1 \over 16}
+{T_2}
-{T_3  \over  8}
-T_4+2T_5-T_6 +2T_7
\eqn$$
vanishes (or rather is a total divergence)
 being proportional to the Euler form.

In order to calculate the appropriate $N=8$ counterterm we evaluate the
(on-shell) amplitude and we find it factorises in the following way:
$$
M^{N=8,D=8}
={1\over \epsilon} \times
\kapss { i \over (4\pi )^4}\; { 1 \over 2}  K_1 \times K_1
\eqn
$$
where
$$
\eqalign{
K_1 &=  tu (\eps_1 \cdot \eps_2) (\eps_3 \cdot \eps_4)
\cr
\null\hskip 0.5 truecm 
&+2 (\eps_1 \cdot \eps_2)
\biggl( t(\eps_3\cdot k_1 \eps_4\cdot k_2)
+u(\eps_3  \cdot k_2   \eps_4\cdot k_1 )
\biggr) 
\cr
&+ \hbox{ cyclic terms} 
\cr}
\eqn
$$
The counterterm necessary to cancel this infinity is,
\def\NORMFACT#1{ {1 \over \eps } \left( {\kappa \over 2} \right)^{4}
{ i \over (4\pi)^{#1} }}
$$
\eqalign{
\NORMFACT{4} 
 {1 \over 4 }
\Bigl[
&-{ T_1 \over 16}
+{T_2}
-{T_3  \over  8}
%\cr
%&
-0.T_4+2T_5-T_6 -2T_7
 \Bigr]
\cr}
\eqn
$$

The other case is $N=4$ supergravity. By this we mean the type~I supergravity
in D=10 and its dimensional descendants. There is of course both a matter multiplet and a gravity multiplet (which contains the graviton.) The $N=8$ multiple
is a sum of these so only one is independent from the $N=8$ case. For the
graviton amplitude with states in the matter multiplet circulating in the loop
the infinity is  
$$
M^{N=4,D=8}
=   { 1 \over \eps}  \times  \kapss { i \over (4\pi )^4}
\times { 1 \over 720}
K_1 \times K_2
\eqn$$
where
$$
\eqalign{
K_2&
=
-\eps_1\cdot \eps_2 \eps_3 \cdot \eps_4
(3t^2+5tu+3u^2)+\cdots
\cr
+&
2  \eps_1\cdot \eps_2 \Bigl(
3s \eps_3 \cdot k_4 \eps_4 \cdot k_3
+t \eps_3 \cdot k_1 \eps_4 \cdot k_2
+u \eps_3 \cdot k_2 \eps_4 \cdot k_1  \Bigr) +\cdots
\cr
-&
12(
k_2\cdot \eps_1
k_1 \cdot\eps_2
k_4\cdot\eps_3
k_3\cdot\eps_4
+
k_3\cdot\eps_1  k_4\cdot\eps_2  k_1\cdot\eps_3  k_2\cdot\eps_4
\cr
&+ k_4\cdot\eps_1  k_3\cdot\eps_2  k_2\cdot\eps_3  k_1\cdot\eps_4 ) \cr}
\eqn
$$
where $+\cdots$ indicates the necessary terms we must add.
We have organised $K_2$ according to the number of $\eps_i\cdot
\eps_j$.  
The counterterms necessary to cancel this are
$$
\eqalign{- \NORMFACT{4}
{1 \over 11520}
\Bigl( -3 T_1 & +24 T_2 -6 T_3
+{4 T_4 }\cr
&
+0.T_5 +0.T_6 +32 T_7
\Bigr)
\cr}
\eqn
$$
We can relate this also to specific tensors contracted against $R^4$.
The tensor $t_8$ can be split into two pieces  
$t_{(12)}$ and $t_{(48)}$,
$t_8={ 1 \over 2 } \Bigl(t_{(12)}+t_{(48)} \Bigr)$
each
having the same symmetry properties as $t_8$. The tensors
$t_{(12)}$ and $t_{(48)}$
contain 
$12$ and $48$ quartic monomials in the $\delta$'s respectively
and are the only two tensors 
which have the same symmetry properties of
$t_8$ in eight dimensions~\cite{GSW}.

%%%%%%% @@@@@@@@@@@@@@
\def\sym4#1#2#3#4{(\delta^{#1#3}\delta^{#2#4}-\delta^{#1#4}\delta^{#2#3}) }
$$
\eqalign{
t_{(12)}^{ijklmnpq}=&-
\biggl(
(\delta^{ik}\delta^{jl}-\delta^{il}\delta^{jk})
(\delta^{mp}\delta^{nq}-\delta^{mq}\delta^{np})
\cr
&+\sym4{k}{l}{m}{n}\sym4{p}{q}{i}{j}
\cr
& 
+\sym4{i}{j}{m}{n}\sym4{k}{l}{p}{q}
\biggr)
\cr
t_{(48)}^{ijklmnpq}=&
\biggl( \delta^{jk}\delta^{lm}\delta^{np}\delta^{qi}
+\delta^{jm}\delta^{nk}\delta^{lp}\delta^{qi}
\cr
+\delta^{jm}&\delta^{np}\delta^{qk}\delta^{li}
+[i\leftrightarrow j]+[k\leftrightarrow l]+[m\leftrightarrow n]
\biggr)\cr
}\eqn
$$
where $[i\leftrightarrow j]$ denotes antisymmetrisation with respect to
$i$ and $j$.
From these tensors we can define
$$
\eqalign{
A\,=\,\frac{1}{4}t_{(12)} t_{(12)}\cdot R^4 %\cr
\;\;, \;\;\; \;   &  B\,=\,\frac{1}{4} t_{(12)} t_{(48)}\cdot R^4 \cr
C\,=\,\frac{1}{4} & t_{(48)} t_{(48)}\cdot R^4 \cr
}
\eqn
$$
where the $\cdot$ denotes the usual contraction of the upper and lower
eight indices.

\noindent ( We can also express these tensor contractions as traces
\cite{Tsey}.)
$
t_{8} t_{(12)}\cdot R^4= 48t_8 \Tr ( R^4 )$
etc )
%
%t_{8}t_{(48)}\cdot R^4= -12t_8  \Tr( R^2)\Tr( R^2)
%\cr}
%\eqn
%$$

In terms of these combinations the $N=8$ counterterm of the type
 $t_8t_8 R^4$ is just
$$
{1 \over 768} \Bigl( A+2B+C \Bigr)
\eqn
$$
and the $N=4$ matter contributions is proportional to 
$\Bigl( 2A+C \Bigr)$.

We have obtained very similar results for $D=10$.  In $D=10$ the $N=8$
supergravity amplitude vanishes (onshell) but the two $N=4$ components
do not.  For these the counterterm also factorises in the form
$
\sim K_1 \times L_i
$
where the tensors $L_i$ contain two more powers of momenta than the $K_i$. 
 
\section{Relationships between gravity and Yang-Mills}

Analysing the structure in amplitudes can reveal strong parallels
between gravity and Yang-Mills calculation.  In many ways the gravity
results appear as the ``square'' Yang-Mills.  In fact, calculations in
$N=4$ Yang-Mills have been used as an initial step in the $N=8$
calculation~\cite{BRY}. To extension to $N=8$ involved to some extend
repeating the calculation whilst squaring the algebra.

Relationships between the tree amplitudes of Yang-Mills were obtained
from  the low energy limit of string theory by Kawai 
Lewellen and Tye~\cite{KLT}. 
They found a series of algebraic
relationships between the two sets of tree amplitudes. At four point
$$
M_4^{\rm tree} (1,2,3,4) =
     - i s_{12} A_4^{\rm tree} (1,2,3,4) \, A_4^{\rm tree}(1,2,4,3)\,,
\eqn
$$
where $M_4$ is the gravity amplitude and $A_4$ is the Yang-Mills
colour ordered tree amplitude.
The KLT relationships have proved extremely useful, however the trees can also be rearranged as in fig.2
where we have rearranged the amplitudes to be in the form
$$
\hbox{kinematic polynomial} \times \hbox{ pole structure}
$$
\EPSFIGURE[r]{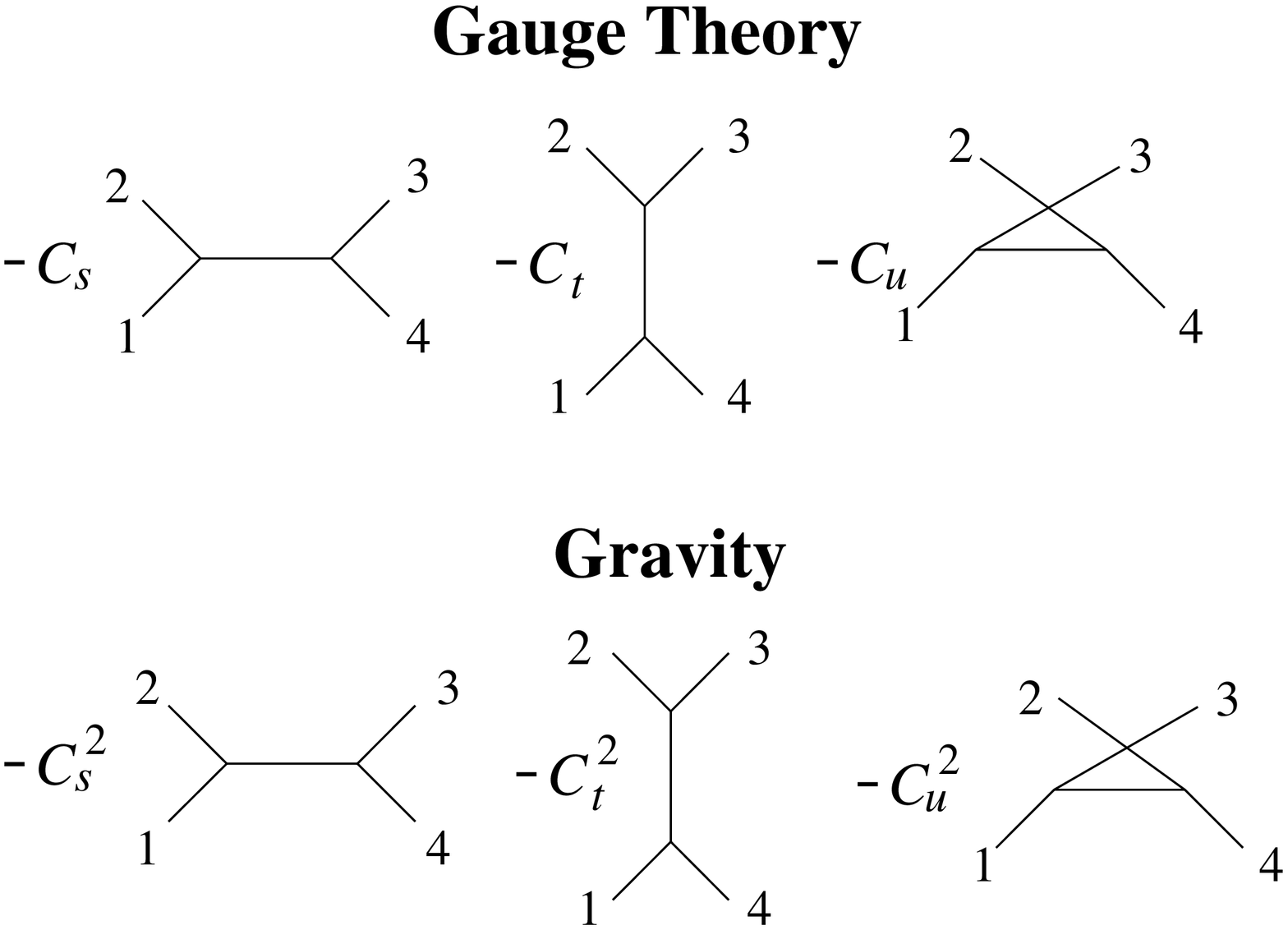,width=4.3truecm}{The tree amplitudes can be
arranged to display a simple squaring relationship between the
Yang-Mills and Gravity Cases}

%\end{document}
This can be done so that the relationship between gravity and
Yang-Mills is clear~\cite{BernGrant}- we keep the pole structure and square the
multiplying polynomial. 
Note that there is some freedom in this since we can
more terms between the different coefficients to some extent. 
\EPSFIGURE[r]{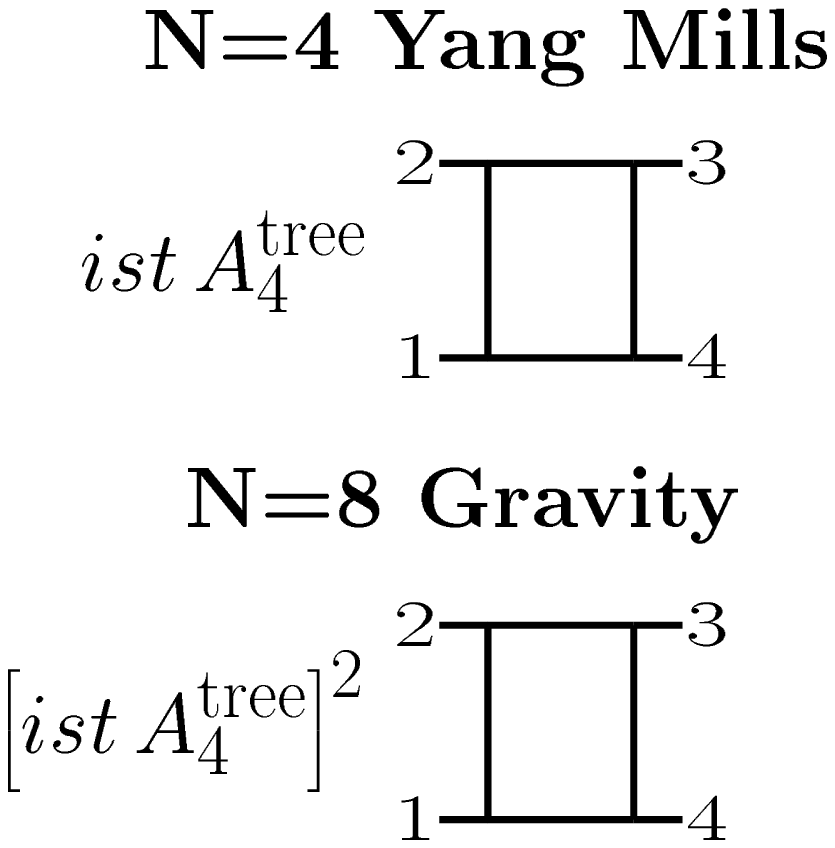,width=3.0 truecm}{One loop relations}

When we consider the one-loop amplitudes we find a very similar
relationship.  The amplitude can be written as a coefficient times an
integral function.  For the case of $N=8$ supergravity the comparison
to the $N=4$ super Yang-Mills is shown in fig.~3 As we can see again we
find that the coefficient of the integral is squared.

For the two loop case we find the situation as shown in fig.4.
(Overall factors of $stA^{\rm tree}$ and $[stA^{\rm tree}]^2$ have
been ommitted for clarity.)  Whilst calculating the supergravity
amplitude this relationship was postulated which allowed an anasatz
for the supergravity amplitude to be quickly made. With a specific
ansatz checking the cuts was relatively straightforward.

\EPSFIGURE[l]{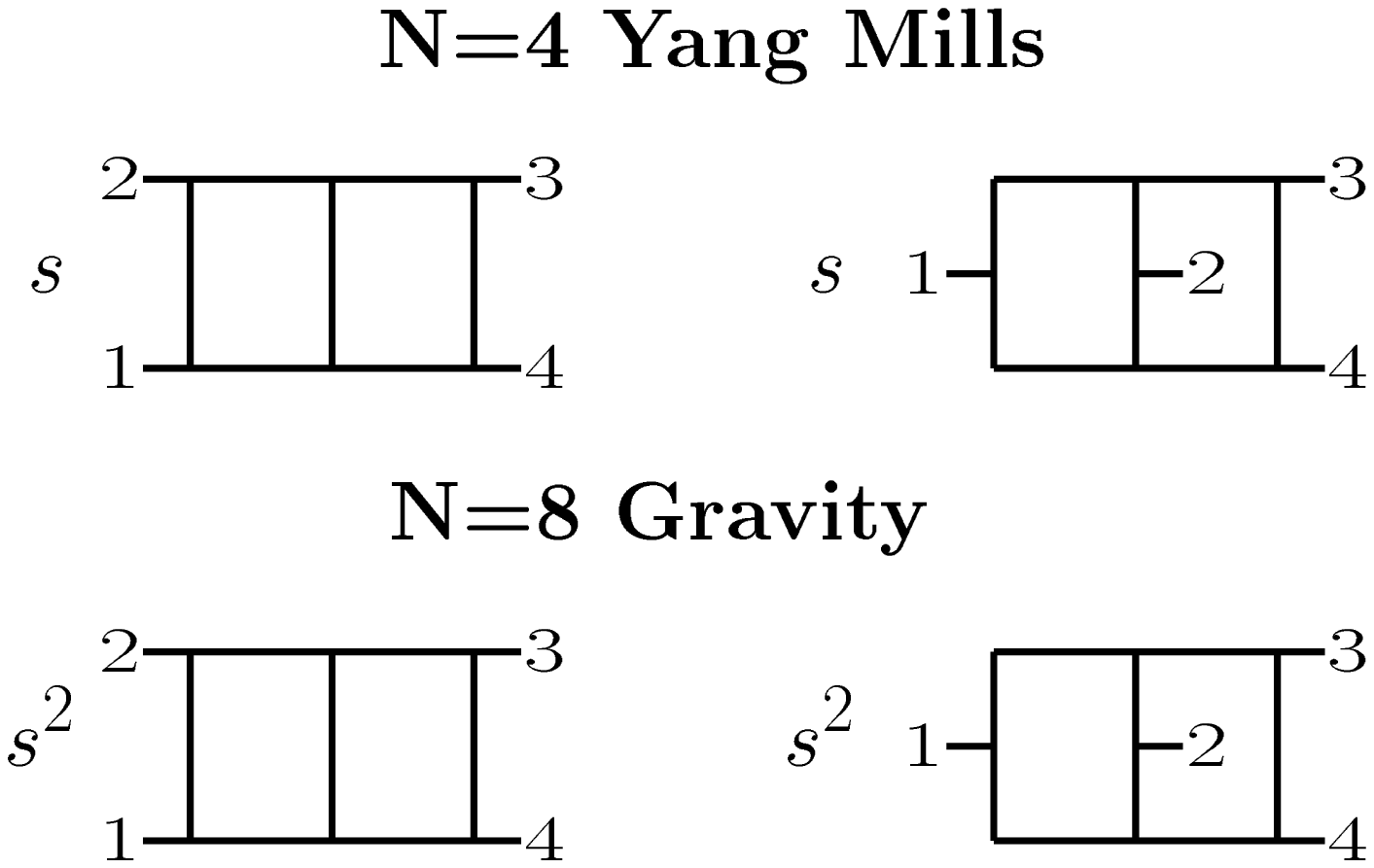,width=3.3 truecm}{A simple relationship between 
the amplitudes of maximal SUSY
and SUGRA persists to two loops}

We suspect the relationships between the perturbative $S$-matrices of
gravity and gauge theories is rather a deep one although our
understanding of it is limited at present.

\vskip 2.0 truecm

%\end{document}


\begin{thebibliography}{99}



\bibitem{Cutting}
L.D.\ Landau, \npb{13}{1959}{181};\\
S. Mandelstam, \pr{112}{1958}{1344}; \pr{115}{1959}{1741}; 
R.E.\ Cutkosky, \jmp{1}{1960}{429};
Z. Bern, L. Dixon, D.C. Dunbar and D.A. Kosower,
\npb{425}{1994}{217}  hep-ph/9403226;
\npb{435}{1995}{59} hep-ph/9409265. 


\bibitem{StringBased}
Z. Bern and D.A.\ Kosower, \npb{379}{1992}{451};
Z. Bern, D.C. Dunbar and T. Shimada,
\plb{312}{1993}{277}  hep-th/9307001;
D.C. Dunbar and P.S. Norridge,
\npb{433}{1995}{181} hep-th/9408014;
Z. Bern and  D.C. Dunbar, 
\npb{379}{1992}{562}
Z.~Bern, \plb{296}{1992}{85};


\bibitem{BCM}
Z. Bern and A.G.\ Morgan, 
\npb{467}{1996}{479} hep-ph/9511336;
Z. Bern, G. Chalmers, L. Dixon and D.A.\ Kosower,
\prl{72}{1994}{2134} hep-ph/9312333;\\
Z.\ Bern and G.\ Chalmers,
\npb{447}{1995}{465} hep-ph/9503236.



%+% 2 refs
\bibitem{ExtendedSugra}
E.\ Cremmer and B.\ Julia,
\plb{80}{1978}{48});
\npb{159}{1979}{141}.


\bibitem{CJS}
E.\ Cremmer, B.\ Julia and J.\ Scherk, \plb{76}{1978}{409}.


\bibitem{GSB}
 M.B.\ Green, J.H.\ Schwarz and L.\ Brink, \npb{198}{1982}{474}.


\bibitem{BDDPR}
Z. Bern, L. Dixon, D.C. Dunbar, M.\ Perelstein and J.S.\ Rozowsky,
\npb{530}{1998}{401} hep-th/9802162;
Class.\ Quant.\ Grav.\  {\bf 17} (2000) 979
hep-th/9911194;
Z.~Bern, L.~Dixon, D.~C.~Dunbar, A.~K.~Grant, M.~Perelstein and J.~S.~Rozowsky,
%``On perturbative gravity and gauge theory,''
Nucl.\ Phys.\ Proc.\ Suppl.\  {\bf 88} (2000) 194
hep-th/0002078.


\bibitem{DS}
S. Deser and D.\ Seminara,
\prl{82}{1999}{2435}, hep-th/9812136.


\bibitem{GSW}
M.B.\ Green, J.H.\ Schwarz and E. Witten,
{\it Superstring Theory} (Cambridge University Press, 1987);

\bibitem{R4}
M.B.~Green, M.~Gutperle and P.~Vanhove,
\plb{409}{1997}{177}, hep-th/9706175;
J.G.\ Russo and  A.A.\ Tseytlin,
\npb{508}{1997}{245}, hep-th/9707134.


\bibitem{BelRob}
I. Robinson, unpublished;\\
L.\ Bel, Acad. Sci. Paris, Comptes Rend. 247:1094 (1958)
and 248:1297 (1959).


\bibitem{HST}
P.S. Howe, K.S. Stelle and P.K. Townsend, 
\npb{236}{1984}{125}.

\bibitem{HoweStelle}
P.S. Howe and K.S. Stelle, 
\plb{137}{1984}{175}
\ijmp{4}{1989}{1871}.

\bibitem{DJST}
D.C.\ Dunbar, B.\ Julia, D.\ Seminara and
M.\ Trigiante, 
\JHEP{\bf 0001}{2000}{046} 
hep-th/9911158. 



\bibitem{Fulling}
S.A.\ Fulling, R.C.\ King, B.G.\ Wybourne and C.J.\ Cummins
\cqg{9}{1992}{1151}.


\bibitem{Tsey}
A.\ Tseytlin, \npb{467}{1996}{383};


\bibitem{BRY}
Z. Bern, J.S. Rozowsky and B. Yan, 
\plb{401}{1997}{273}
hep-ph/9702424.

\bibitem{KLT}
H. Kawai, D.C. Lewellen and S.-H.H. Tye, Nucl.\ Phys.\ B269:1 (1986).

\bibitem{BernGrant}
Z.~Bern and A.K.~Grant,
\plb{457}{1999}{23}, hep-th/9904026;


\bibitem{DNb}
D.C. Dunbar and P.S. Norridge, 
\cqg{14}{1997}{351}
hep-th/9512084;\\
P.S. Norridge,  
\plb{387}{1996}{701}  hep-th/9606067.


\end{thebibliography}
\end{document}